\documentclass[a4paper]{jpconf}
\usepackage{graphicx}
\usepackage{amsmath}
\begin{document}
\title{Gravitation, and quantum theory, as emergent phenomena}

\author{Tejinder P. Singh$^{a,b}$}

\address{{}$^{a}$Inter-University Centre for Astronomy and Astrophysics, Post Bag 4, Ganeshkhind, Pune 411007, India \\
{}$^{b}$Tata Institute of Fundamental Research, Homi Bhabha Road, Mumbai 400005, India}

\ead{tejinder.singh@iucaa.in, tpsingh@tifr.res.in
}

\begin{abstract}
There must exist a reformulation of quantum field theory, even at low energies, which does not depend on classical time. The octonionic theory proposes such a reformulation, leading to a pre-quantum pre-spacetime theory. The ingredients for constructing such a theory, which is also a unification of the standard model with gravitation, are : (i) the pre-quantum theory of trace dynamics – a matrix-valued Lagrangian dynamics, (ii) the spectral action principle of non-commutative geometry, (iii) the number system known as the octonions, for constructing a non-commutative manifold and for defining elementary particles via Clifford algebras, (iv) a Lagrangian with $E_8 \times E_8$ symmetry. The split bioctonions define a sixteen dimensional space (with left-right symmetry) whose geometry (evolving in Connes time) relates to the four known fundamental forces, while predicting two new forces, $SU(3)_{grav}$ and $U(1)_{grav}$. This latter interaction is possibly the theoretical origin of MOND. Coupling constants of the standard model result from left-right symmetry breaking, and their values are theoretically determined by the characteristic equation of the exceptional Jordan algebra of the octonions. The quantum-to-classical transition, precipitated by the entanglement of a critical number of fermions, is responsible for the emergence of classical spacetime, and also for the familiar formulation of quantum theory on a spacetime background.

\end{abstract}

\section{Introduction: Quantum theory without classical time, as a route to quantum gravity and unification}

\smallskip

In the words of Edward Witten \cite{Witten1986},

\smallskip

\noindent ``If one wants to summarise our knowledge of physics in the briefest possible terms, there are three really fundamental observations: (i) Space-time is a pseudo-Riemannian manifold $M$, endowed with a metric tensor and governed by geometrical laws. (ii) Over $M$ is a vector bundle $X$ with a nonabelian gauge group $G$. (iii) Fermions are sections of $(\hat{S}_{+} \otimes V_R)\oplus(\hat{S}_{-}\otimes V_{\bar{R}})$. $R$ and $\bar{R}$ are not isomorphic; their failure to be isomorphic explains why the light fermions are light and presumably has its origins in a representation difference $\Delta$ in some underlying theory. All of this must be supplemented with the understanding that the geometrical laws obeyed by the metric tensor, the gauge fields, and the fermions are to be interpreted in quantum mechanical terms''  [{\it from the CERN preprint `Physics and Geometry' (1987)}]. 

We attempt to bring the above-quoted three observations into one unified framework, in the octonionic theory \cite{SinghTD, Singh2021} summarised in the present article. This unification is based on Adler's theory of trace dynamics \cite{Adler04} and on the use of the number system of octonions as coordinate systems. The resulting theoretical framework has some commonalities with string theory (extended objects, $E_8\times E_8$ symmetry) but also significant differences (octonions as coordinates, no compactification, trace dynamics instead of quantum dynamics). These differences help overcome the challenges to  string theory as a theory of unification - we do not have multiverses, landscape or swampland; but rather the  emergence of the standard model and of spacetime geometry, from an underlying higher dimensional geometric theory possessing $E_8 \times E_8$ symmetry.

The central premise for this new approach to unification is the following. The classical spacetime manifold (labelled by real numbers) exists if and only if the universe is dominated by classical bodies (planets, stars, galaxies). Therefore, the use of a classical spacetime background in quantum field theory is an approximation, because the classical objects which facilitate such a background to exist are themselves a limiting case of quantum systems. There must hence exist a reformulation of quantum field theory, which should make no reference to classical spacetime, and such a reformulation must exist at all energy scales, not just at the Planck energy scale. For there is nothing that in principle prevents a low energy universe from being entirely devoid of classical bodies, and even in such a universe we must be able to describe the dynamics of elementary particles. It turns out that a description of this kind, which uses octonions instead of real numbers as coordinates, explains why the standard model (including its 26 free parameters) is what it is, and offers a way to unify it with gravitation.

To reiterate the basic principle, consider that every elementary particle in the universe is in a quantum superposition of two or more position states. The corresponding gravitational fields will also be in a quantum superposition. Consequently, the operational distinguishability of space-time points will be lost, this latter being an implication of the Einstein hole argument, which requires the spacetime manifold to be overlaid by a classical metric. This argument holds true at every energy scale. The loss of classical spacetime hence necessitates that there be a reformulation of quantum field theory without classical spacetime; in particular needed also (and with far-reaching consequences) when we describe  the standard model of particle physics.

We develop such a reformulation using only Planck length $L_p$, Planck time $\tau_p$, and Planck's constant $\hbar$ as the fundamental constants of the theory. Every other dimensionful constant is expressible in terms of these (e.g. Newton's gravitational constant $G_N = L_p^5 / \hbar \tau_p^3$) and every dimensionless constant (made with or without using these three constants) must be derivable from first principles. We have traded $\hbar$ for Planck energy, with the latter now assumed a derived quantity ($=\hbar/\tau_p$). When every physical subsystem in a chosen system has an action of the order $\hbar$, the point structure of spacetime is lost, irrespective of energy scale. In addition, if the length scale / time scale of interest is Planck length / Planck time, the energy scale is the Planck scale, and quantum gravitational effects become significant. Clearly, the point structure of spacetime can in principle be lost even if quantum gravity effects are not significant. The familiar formulation of quantum dynamics on a classical spacetime background is possible only when the universe is dominated by physical subsystems each of which has an action much larger than $\hbar$.
\begin{figure}[h]
\centering
\includegraphics[width=15cm]{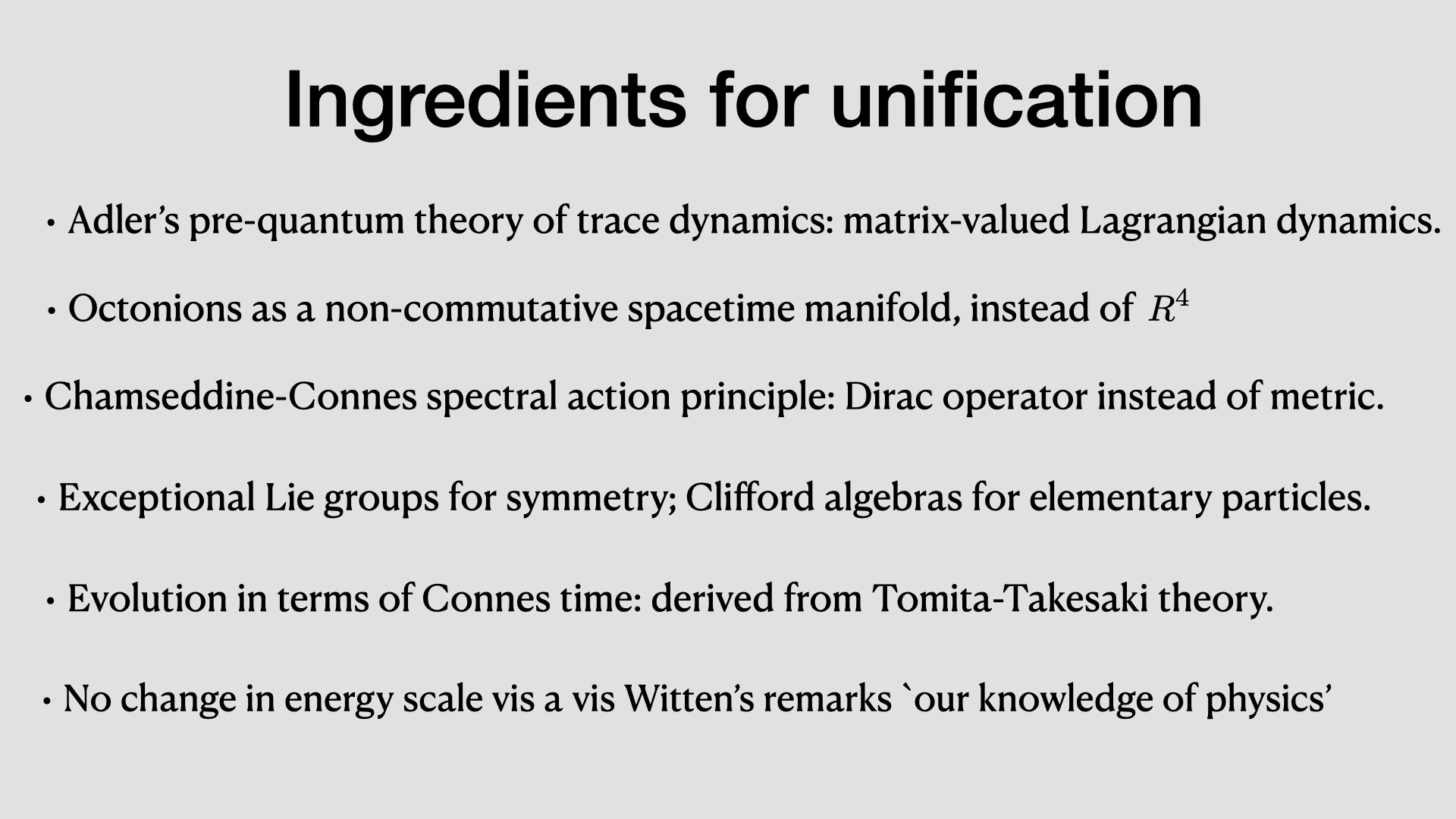}
\caption{The ingredients for unification in the octonionic theory}
\end{figure}

Fig. 1 lists the various ingredients which go into the construction of the proposed theory of unification. 
\begin{figure}[h]
\centering
\includegraphics[width=15cm]{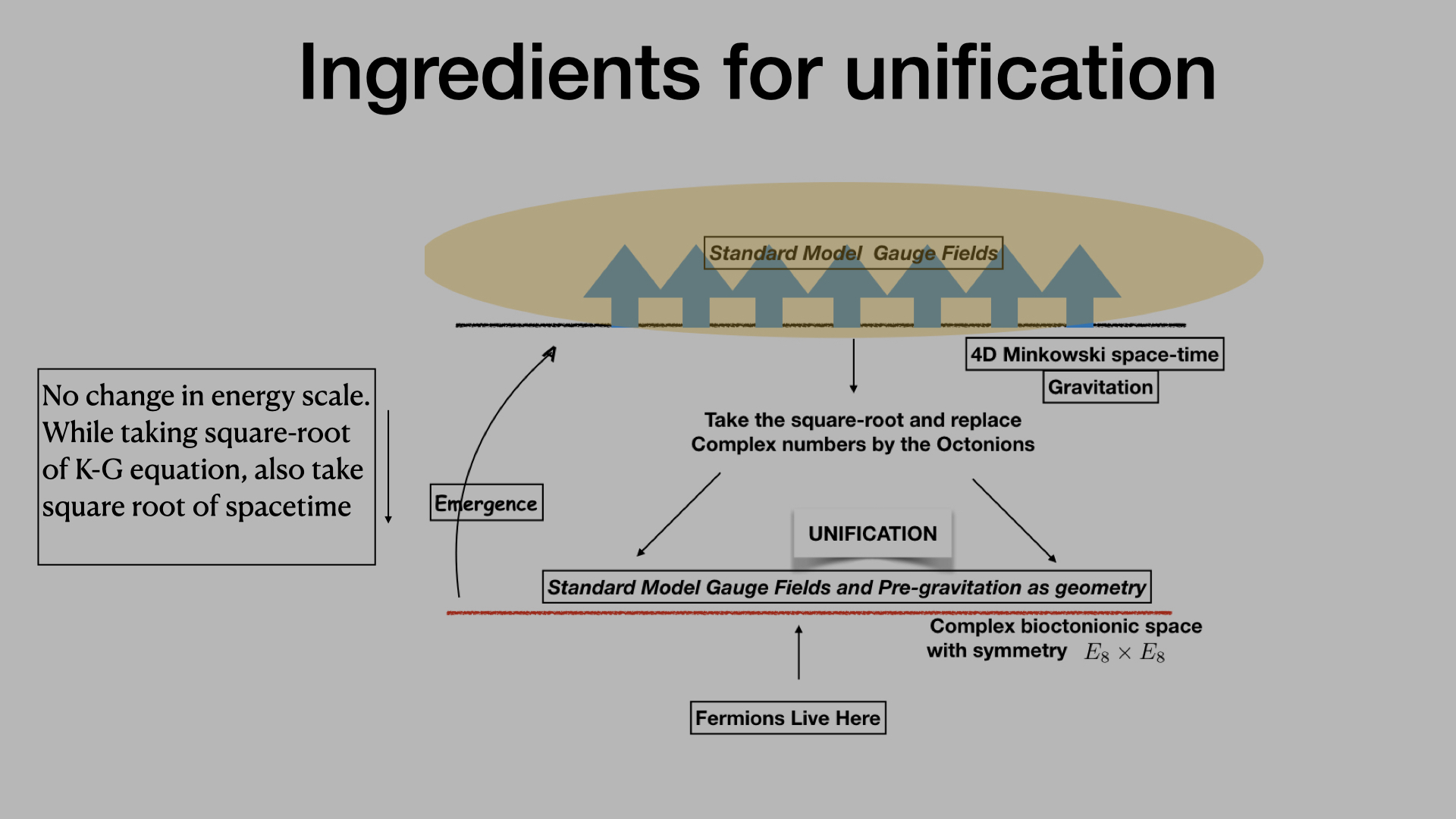}
\caption{Ingredients for unification: taking the square root of Minkowski spacetime}
\end{figure}
Fig. 2 highlights another aspect of this unification. Standard model gauge fields and fermions are assumed to live on a 4D spacetime curved by gravitation. When we take the square-root of the Klein-Gordon equation to write the Dirac equation, let us also take the square root of the Minkowski line element, i.e. describe the spacetime in spinorial language, as a twistor space, using complex numbers.
If we were to replace complex numbers by the quaternions, the Dirac operator can be shown to be the gradient operator on the quaternionic space. Next, we replace the quaternions by the octonions (more precisely, complex split bioctonions). Standard model gauge fields and pre-gravitation are found to describe the geometry of this bioctonionic space ($E_8 \times E_8$ symmetry), and fermions in reality live in this space, not in spacetime. When very many degrees of freedom get entangled so that several physical subsystems each have action much larger than $\hbar$, classical spacetime emerges (`Emergence' in Fig. 2). Those subsystems which still have action order $\hbar$ are quantum in nature, and should strictly be described on bioctonionic space, but can be described, to  a good approximation, on the emergent 4D spacetime background accompanied by the aforesaid vector bundle.

In the following sections we describe these ingredients, and the proposed theory, in some detail. 

\section{From Newtonian dynamics to trace dynamics: quantum theory as an emergent phenomenon}
Consider that we wish to define Newtonian dynamics using, not real numbers, but matrices. For instance, given the action 
\begin{equation}
S = \sum_i\;  \int d\tau \; \frac{1}{2}m_i\left(\frac{dq_i}{d\tau}\right)^2
\end{equation}
for a collection of point particles with configuration variables labelled by real numbers $q_i$, we replace real numbers by matrices, $q_i \rightarrow {\bf q_i}$. The Lagrangian becomes a matrix polynomial, and the new Lagrangian for this matrix dynamics is defined as the trace of the matrix polynomial:
\begin{equation}
S = \sum_i\;  \int d\tau \; Tr\left[ \frac{1}{2}\frac{L_{p}^2}{L^2}\left(\frac{d{\bf q}_i}{d\tau}\right)^2 \right]
\end{equation}
[Keeping in view what lies ahead, the mass parameter $m$ has been replaced by (square of a) length parameter $L$ (measured in units of Planck length $L_p$)]. Lagrange equations of motion are derived by varying the trace Lagrangian with respect to the matrix variables, and an equivalent Hamiltonian formulation and a phase space dynamics is also developed. Hamilton's equations of motion are precursors of the Heisenberg equations of motion of quantum theory; the latter are emergent.

This is the theory of trace dynamics, developed by Adler and collaborators \cite{Adler1, Adler2}. The matrices (equivalently operators) in this matrix-valued Lagrangian dynamics have the same status as in Heisenberg's matrix mechanics; their eigenvalues are the values that the corresponding classical dynamical variables take during evolution. Trace dynamics is more general than quantum theory, because it possesses an additional conserved Noether charge, because of the invariance of the trace Hamiltonian under global unitary transformations. This charge, denoted $\tilde{C}$ and  known as the Adler-Millard charge, is given in terms of canonical configuration variables and their corresponding momenta, as
\begin{equation}
\tilde{C} = \sum_i \; [{\bf q}_B, {\bf p}_B] -\left\{{\bf q}_F,{\bf p}_F\right\}
\end{equation}
The matrices have Grassmann numbers as their entries (analogous to the case in quantum field theory); the matrices $q_B$ are made of even grade Grassmann numbers and known as bosonic matrices; whereas the $q_F$ made of odd grade Grassmann numbers are known as fermionic matrices. This conserved charge has the dimensions of action, and is what make trace dynamics  into a 
pre-quantum theory, from which quantum field theory is emergent. The continuum field theory generalisation of trace dynamics is achieved by considering a classical field as a collection of point particles (one per space-time point), generalising each such point particle to a matrix, and then integrating the trace Lagrangian over space-time volume so as to obtain the action. Furthermore, one can start from a Lorentz-invariant classical dynamics and construct its generalisation to a relativistic trace dynamics.

It is assumed that trace dynamics holds at some time scale resolution not accessed by current 
laboratory experiments, say Planck time $\tau_{p}$. We then ask what is the emergent coarse-grained dynamics, if the system is observed not at Planck time resolution, but at some lower resolution $\tau\gg \tau_p$? The standard techniques of statistical thermodynamics are employed to construct a phase space density distribution of the trace dynamical system, whose emergent coarse-grained dynamics is determined by maximising the Boltzmann entropy subject to constraints representing conserved quantities. It is shown that at thermodynamic equilibrium the Adler-Millard charge is equipartitioned over all degrees of freedom so that the canonical average of each commutator $[q_B, p_B]$ and each 
anti-commutator $\{q_F, p_F\}$ is assumed to be equal to $i\hbar$. This is how the quantum commutation relations emerge from the underlying trace dynamics. Also, in this emergent thermodynamic equilibrium, the canonically averaged Hamilton's equations of motion become Heisenberg's equations of motion of quantum theory. Identification of canonical averages of functions of dynamical variables (in their ground state) with Wightman functions in relativistic quantum mechanics enables the transition from trace dynamics to quantum  field theory. Quantum theory is thus shown to be an emergent (equilibrium) thermodynamic phenomenon.

At equilibrium, the Adler-Millard charge is anti-self-adjoint, and the Hamiltonian of the theory is self-adjoint. Statistical fluctuations in this charge, when significant, can drive the quantum system away from equilibrium (the charge is no longer equipartitioned). If these fluctuations are themselves dominantly self-adjoint, the Hamiltonian of the theory picks up an anti-self-adjoint component, which gets amplified if a large number of degrees of freedom are entangled with each other. This drives the system to classicality, via a Ghirardi-Rimini-Weber type of spontaneous collapse process. Thus, macroscopic classical systems are far from equilibrium emergent states in trace dynamics. 

If the fluctuations in the Adler-Millard charge are dominantly anti-self-adjoint, the Hamiltonian of the theory continues to be self-adjoint and the system can be said to be in a quantum non-equilibrium, perhaps analogous to the quantum non-equilibrium talked of in Bohmian mechanics. The quantum commutation relations no longer hold. As a consequence, such trace dynamical systems violate the Tsirelson bound (obeyed by quantum systems) of the CHSH inequality. It has long been a puzzle as to why the Popescu-Rohrlich bound of $4$ permitted by relativistic causality is higher than the Tsirelson bound of $2\sqrt{2}$ obeyed by quantum mechanical systems. We now know the answer: trace dynamics, being more general than quantum theory, permits supra-quantum non-local correlations when the Adler-Millard charge is not equipartitioned and is dominantly anti-self-adjoint \cite{Ahmed}. This situation is exhibited in Fig. 3 and the experimental search for such supra-quantum correlations is of great interest: a confirmation will be a conclusive signature that quantum theory is approximate, not exact. In fact, the quantum state is an attractor, being the equilibrium state, to which classical systems as well as supra-quantum systems evolve.
\begin{figure}[h]
\centering
\includegraphics[width=15cm]{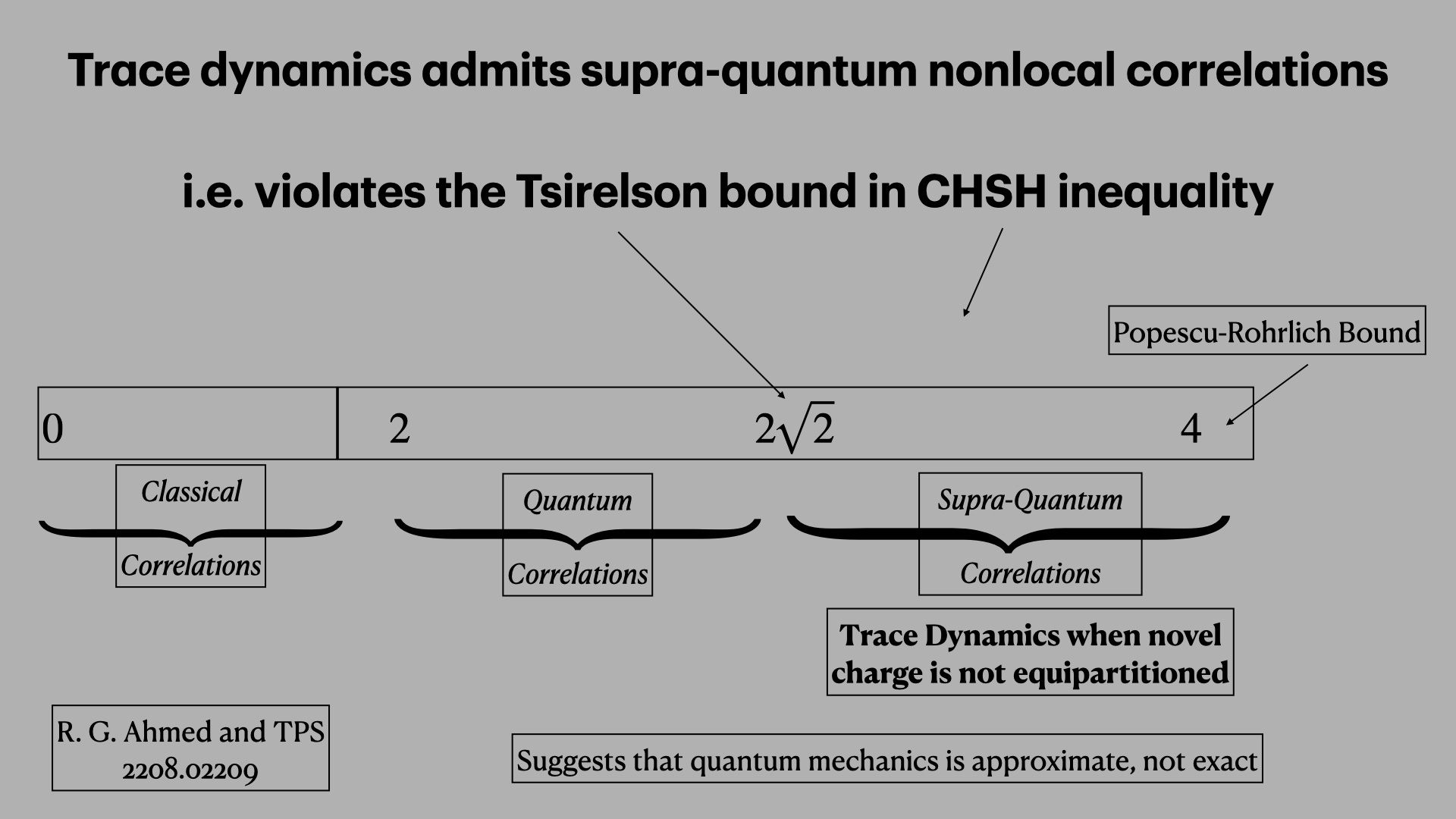}
\caption{Trace dynamics violates the Tsirelson bound in the CHSH inequality}
\end{figure}Since the Adler-Millard charge has dimensions of action, its conjugate variable must be dimensionless. This conjugate variable is a time parameter $\tau$ measured in units of Planck time $\tau_p$; however this is not the time coordinate (this latter is usually denoted $t$) of the spacetime manifold of special relativity. Rather the time $\tau$ is the Connes-Tomita-Takesaki time parameter (to be discussed in the next section) which is a unique feature of non-commutative geometry resulting from the Tomita-Takesaki theory. The conjugate of coordinate time is energy; the conjugate of Connes time is the Adler-Millard charge. When $\tau/\tau_p\gg 1$, the Adler-Millard charge is equipartitioned; and when $\tau/\tau_p\sim 1$, the charge is not equipartitioned and supra-quantum non-local quantum correlations arise. It might hence seem that such correlations can only be realised at Planck time resolution. However, in the octonionic theory presently under review, an `octonionic' inflation in the very early universe resets the Planck energy scale to the TeV scale; hence it might be possible to experimentally detect violation of the Tsirelson bound by performing Bell type experiments at around the TeV scale. Also, it appears that trace dynamics needs non-commutative geometry for a consistent interpretation of the Adler-Millard charge. As originally formulated, trace dynamics assumes a classical spacetime manifold, which could be flat, or endowed with classical gravitation.

We generalise trace dynamics to remove the classical spacetime manifold, by using the non-commuting octonions as coordinates instead. Also, a matrix-valued description of gravitation is developed, by using the spectral action principle of Connes' non-commutative geometry. This leads to a pre-quantum,
 pre-spacetime theory from which classical spacetime geometry and quantum theory are emergent. Also, trace dynamics does not specify the fundamental Lagrangian of the universe. In the octonionic theory, we propose a trace dynamics Lagrangian with $E_8 \times E_8$ symmetry, which unifies 
 pre-gravitation with the standard model.
 
 \section{Choosing the Lagrangian: the spectral action principle}
 In order to decide as to how gravitation should be incorporated in trace dynamics, we appeal to the celebrated spectral action principle \cite{Chams}, used in particular to make the transition from Riemannian geometry to non-commutative geometry.  According to this principle, the Einstein-Hilbert action can be expressed in terms of the eigenvalues of the Dirac operator $D$ on a Riemannian geometry, via a truncated heat kernel expansion in powers of $L_p^{-2}$:
 \begin{equation}
 Tr [L_P^2 D^2] \sim L_p^{-2}\int d^4x\; \sqrt{g}\; R + {\cal O}(L_P^{0}) =\sum_{i}L_p^2\; \lambda_i^2 \end{equation}
It has also been shown that these Dirac eigenvalues can be the dynamical observables of general relativity \cite{Landi}, in place of the metric. In the spirit of trace dynamics, every eigenvalue $\lambda_i$ is raised to the status of a bosonic matrix/operator $\hat{\lambda}_i\equiv \dot{q}_{Bi}$, this being the very Dirac operator $D$ of which it is an eigenvalue. Thus in generalised trace dynamics (a pre-quantum, pre-spacetime theory) we have a collection of `atoms of space-time', as many atoms as there were Dirac eigenvalues, each atom being associated with a copy of the Dirac operator: $\dot{q}_{Bi}\equiv LD$, where $L$ is a newly introduced length parameter which characterises a space-time atom. The Einstein-Hilbert action $\sum_i\; L_p^2\;  \lambda_i^2$ transits to $\sum_i Tr \; (L_p^2/L^2)\; [\dot{q}_{Bi}^2]$. Since the Dirac eigenvalues have been made operators, space-time is lost, and this is simultaneously a transition to trace dynamics and to non-commutative geometry, hence showing the deep connection between the new geometry and the new dynamics. But with one caveat: what was earlier the (dimensionless) action is now the dimensionless (trace) Lagrangian; the integral over time, which will make it into a trace dynamics action, is missing! Here, Connes time parameter $\tau$, a unique feature of non-commutative geometry, comes to our rescue. The trace dynamics action for atoms of space-time matter, scaled with respect to Planck's constant $\hbar$, is given by
\begin{equation} 
\frac{S}{\hbar} = \int \frac{d\tau}{\tau_p} \; \sum_i Tr \; (L_p^2/L^2)\; [\dot{q}_{Bi}^2]
\end{equation}
The spectral action principle has been shown to hold also when Yang-Mills gauge fields are present, besides gravitation; with the gauge field $A^\mu$ introduced in the conventional manner of modifying the Dirac operator: $D\rightarrow D + A^\mu$. The eigenvalues of the gauge potential are raised to the status of matrices, and these matrices are identified with the configuration variables $q_{Bi}$ of the corresponding Dirac operators  $\dot{q}_{Bi}$. Matter (fermionic) degrees of freedom $q_{Fi}$ and $\dot{q}_{Fi}$ are introduced such that the eigenvalues of these matrices relate to the classical relativistic action of point particles (mass term as well as currents related to gauge fields). 

Henceforth, we will focus on just one STM (space-time-matter) atom and try to understand its properties; leaving for later the question of interaction between several atoms. If we keep only the dotted terms, we have the action for gravitation coupled to (as we shall see, right handed) fermions:
\begin{equation}
\frac{S}{\hbar}= \int \frac{d\tau}{\tau_{p}}\; Tr \bigg\{ \frac{L_p^2}{L^2}\left[\dot{q}_B^\dagger +  \frac{L_p^2}{L^2}\beta_1 \dot{q}_F^\dagger \right ] \times \left[\dot{q}_B + \frac{L_p^2}{L^2} \beta_2 \dot{q}_F\right] \bigg\}
\end{equation}
$\beta_1$ and $\beta_2$ are constant Grassmann elements introduced to make the Lagrangian bosonic.
When the undotted terms are also included, we get the action also for the Yang-Mills fields and 
left-handed fermions, all put together this  defines an `atom of space-time-matter', or an {\it aikyon}. 
\begin{equation}
\frac{S}{\hbar}= \int \frac{d\tau}{\tau_{p}}\; Tr \bigg\{ \frac{L_p^2}{L^2}\left[\left(\dot{q}_B^\dagger + i\frac{\alpha}{L} q_B^\dagger\right) + \frac{L_p^2}{L^2}\beta_1\left( \dot{q}_F^\dagger + i\frac{\alpha}{L}q_F^\dagger\right)\right ] \times \left[\left(\dot{q}_B + i\frac{\alpha}{L}q_B\right) + \frac{L_p^2}{L^2} \beta_2 \left(\dot{q}_F+ i\frac{\alpha}{L}q_F\right)\right] \bigg\}
 \label{aikyacn2}
 \end{equation}
 An aikyon is an elementary particle (say an electron) along with all the bosonic fields it produces. (The word aikyon derives from the Sanskrit word {\it aikya} meaning `oneness'.) 
 This is the fundamental action principle for the octonionic theory. The claim is that when these matrices are defined on the space of split bioctonions, and the Lagrangian has $E_8 \times E_8$ symmetry, this action principle describes (after left-right symmetry breaking) standard model gauge fields and chiral fermions and the Higgs, as also an additional Higgs and (a generalisation of) Einstein's general relativity which now  includes also  an $SU(3)_{grav}$ interaction and an $U(1)_{grav}$ interaction, with the latter showing strong evidence for being the origin of (relativistic) MOND \cite{Jose}. General relativity itself is inferred as the right-handed counterpart of the weak force, both being broken symmetries! The constant $\alpha$ is the Yang-Mills coupling constant, originating from $E_8\times E_8$ - it arises as a result of left-right symmetry breaking which separates the unified dynamical variable $\dot{Q}_B$ into its gravitational and gauge sectors: $\dot{Q}_B\equiv \dot{q}_B + (\alpha/L) q_B$.
 
 The Hamiltonian of the theory is not self-adjoint. Assuming that the theory holds at Planck (Connes) time resolution, and assuming that the anti-self-adjoint part of the Hamiltonian is insignificant, coarse graining to a lower time resolution gives the sought for reformulation of quantum theory which does not depend on classical spacetime. This is also a quantum theory of gravity. Evolution continues to be defined through Connes time. If in the underlying theory, sufficiently many degrees of freedom get entangled, the imaginary part of the Hamiltonian becomes important, leading to collapse of superpositions [a deterministic, non-unitary and norm-preserving evolution]. If this system is observed only under a coarse-grained approximation, the outcomes of collapse appear random, while obeying the Born probability rule, and thus offering a theoretical underpinning for models of objective collapse \cite{Kakade}. Collapse is the opposite of raising each eigenvalue to the status of a matrix: spontaneous collapse sends the matrix back to one of its eigenvalues. In this process one also recovers classical space-time and general relativity, as the Dirac operators collapse to (distinct) eigenvalues, and the spectral action principle ensures recovery of the Einstein-Hilbert action coupled to Yang-Mills fields and relativistic point particles, with additional corrections. Those degrees of freedom which are not sufficiently entangled continue to obey the underlying trace dynamics (no spacetime) but can be described to a  good approximation by conventional quantum field theory on a classical spacetime background. In making this approximation, the origins of the standard model of particle physics are lost. We note that this spacetime background has become available only because the universe is dominated by classical objects.
 
 Everything that is said in the previous paragraph is independent of energy scale. When we say that Connes time $\tau$ is being measured at Planck time resolution, it does not imply that the system is being probed at Planck energies. The conjugate variable for Connes time is not energy, but the Adler-Millard charge: at Planck time resolution this charge is not equipartitioned; at lower resolution it is. The variable conjugate to energy is the coordinate time of special relativity - this time does not flow and in fact, calling it time (time being that which has an arrow and distinguishes past from future) seems like a misnomer! When evolution is described through Connes time, the emergent quantum theory is analogous to the Stueckelberg-Hurwitz formulation of relativistic quantum mechanics.
 
So far, we have a matrix-valued Lagrangian dynamics, which is a generalisation of classical 
real-number valued dynamics. We have also made a transition from Riemannian geometry to Connes' non-commutative geometry. What remains is to transit from the real-number valued coordinate system which labels the 4D space-time manifold, and to instead work with the non-commuting numbers known as quaternions and octonions. The dynamical matrices (which replace vectors) have matrix-valued `coordinate' components over the field of quaternions/octonions, instead of over the field of real numbers. We then have a pre-quantum, pre-spacetime dynamics in higher dimensions, which we employ to describe the standard model as well as gravitation, because the (broken) symmetries of 
bi-octonionic space coincide with the ones observed in nature.

\section{Octonions as coordinate systems: a non-commutative manifold}
 At the beginning of this article, we quoted Witten's remarks: ``(i) Space-time is a pseudo-Riemannian manifold $M$, endowed with a metric tensor and governed by geometrical laws. (ii) Over $M$ is a vector bundle $X$ with a non-abelian gauge group $G$." We are going to employ octonionic space to unify this vector bundle and the 4D space-time manifold into a new higher dimensional space. This is done without change of energy scale, at the very energies at which the standard model is formulated at present. The algebra automorphisms of the octonions unify space-time diffeomorphisms and standard model gauge field transformations into one common symmetry ($E_8 \times E_8$). The octonionic coordinate space is defined separately for every atom of space-time-matter, one coordinate copy per atom.
 
There are only four division algebras: reals, complex numbers, quaternions and octonions, denoted $\mathbf{R, C, H, O}$. A quaternion $\mathbf{H}$ 
\begin{equation}
\mathbf{H} = 
a_0 + a_1 \hat i + a_2 \hat j + a_3 \hat k; \quad \hat i^2 = \hat j^2 = \hat k^2 = -1;   \quad \hat i \hat j = \hat k = - \hat j \hat i;   \ \hat j \hat k = \hat i = - \hat k   \hat j;   \ \hat k \hat i = \hat j = - \hat i \hat k
\end{equation}
can be used to define a vector and its rotations in 3D space. A split biquaternion is defined as
\begin{equation}
\mathbf{H} \oplus \omega \mathbf {\tilde{H}} = (a_0 + a_1 \hat i + a_2 \hat j + a_3 \hat k) \oplus \omega (a_0 - a_1 \hat i - a_2 \hat j - a_3 \hat k)
\end{equation}
Here $\omega$ is the split complex number (i.e. $\omega^*=-\omega, \omega^2=1$) made from the imaginary directions of a quaternion. Complexified split biquaternions are key to defining chiral leptons in this theory. Furthermore, the Dirac operator is nothing but the gradient operator on quaternionic space - the gamma matrices present in the Dirac operator when defined on Minkowski spacetime mimic the true nature of spacetime, which is quaternionic and non-commutative. The Lagrangian we have constructed in (\ref{aikyacn2}) is essentially the square of the Dirac operator (squared momentum / kinetic energy) of a free particle.

An octonion is defined as \cite{Baez}
\begin{equation}
\mathbf{O} = a_0 + a_1 {\bf e}_1 + a_2 {\bf e}_2 + a_3 {\bf e}_3 + a_4 {\bf e}_4 + a_5 {\bf e}_5 + a_6 {\bf e}_6 + a_7 {\bf e}_7
\end{equation}
The seven imaginary direction anti-commute, each of them squares to $-1$, and octonionic multiplication obeys the Fano plane rules. A split bioctonion is defined as
\begin{equation}
\begin{split}
\mathbf{O} + \omega \mathbf{\tilde O}= & (a_0 + a_1 {\bf e}_1 + a_2 {\bf e}_2 + a_3 {\bf e}_3 + a_4 {\bf e}_4 + a_5 {\bf e}_5 + a_6 {\bf e}_6 + a_7 {\bf e}_7) \ + \\
& \omega(a_0 - a_1 {\bf e}_1 - a_2 {\bf e}_2 - a_3 {\bf e}_3 - a_4 {\bf e}_4 - a_5 {\bf e}_5 - a_6 {\bf e}_6 - a_7 {\bf e}_7)
\end{split}
\label{spbioc}
\end{equation}
This time the split complex number $\omega$ is made from the imaginary directions of the octonion. Complexified split bioctonions are central to defining chiral quarks and leptons. Whereas split biquaternions are adequate for chiral leptons, the extension to split bioctonions is essential for bringing in chiral quarks: QCD is the geometry of extra spatial dimensions (there being four such extra dimensions). 

Bosons and fermions are defined on split bioctonionic space; for instance
\begin{equation}
Q_B = Q_0 + Q_1 {\bf e}_1 + Q_2 {\bf e}_2 + Q_3 {\bf e}_3 + Q_4 {\bf e}_4 + Q_5 {\bf e}_5 + Q_6 {\bf e}_6 + Q_7 {\bf e}_7
\end{equation}
shows the matrix-valued components $Q_i$ of a bosonic matrix $Q_B$ over octonionic space. In the action (\ref{aikyacn2}) the undotted matrices are defined over octonionic space and dotted matrices over the split part of the bioctonionic space. Keeping this in mind, consider the modulus square of the split bioctonion:
\begin{equation}
\begin{split}
|O + \omega \tilde{O}|^2 & = (\tilde O \ominus \omega O) \times (O \oplus  \omega \tilde O) = \tilde{O}O \oplus \omega \tilde O \tilde O \ominus  \omega O O \ominus O \tilde O   \\
& =  (a_0^2 + a_1^2 + a_2^2 + a_3^2 + a_4^2 + a_5^2 + a_6^2 + a_7^2)\  \oplus \\
&\ \ \ \   \omega (a_0^2 - a_1^2 - a_2^2 - a_3^2 - a_4^2 - a_5^2 - a_6^2 - a_7^2 +Im1) \ \ominus \\
& \ \ \ \ \omega (a_0^2 - a_1^2 - a_2^2 - a_3^2 - a_4^2 - a_5^2 - a_6^2 - a_7^2 + Im2)\  \ominus  \\
&\ \ \ \   \   (a_0^2 + a_1^2 + a_2^2 + a_3^2 + a_4^2 + a_5^2 + a_6^2 +a_7^2) 
\end{split}
\label{thirteen}
\end{equation}
The four expressions in the four lines after the second equality demonstrate the unified presence of the vector bundle (lines one and four, Euclidean line-element) and space-time (lines two and three, Lorentzian line element, with imaginary corrections). Inspecting the bosonic part of the Lagrangian (\ref{aikyacn2}) we see that the two Euclidean elements are for the dotted quadratic term $\dot{q}_B^\dagger \dot{q}_B$ and the undotted term ${q}_B^\dagger q_{B}$ respectively. As has been analysed in \cite{Raj}, and supported by the results in \cite{Kaushik}, the undotted term represents an interaction with $SU(3)$ symmetry that is identified with $SU(3)_{color}$, whereas the dotted term is a new $SU(3)$ symmetry interpreted as $SU(3)_{grav}$. The Lorentzian elements in lines two and three in the above equation are for the mixed terms $\dot{q}_B^\dagger q_B$ and ${q}_B^\dagger  \dot{q}_B$. They represent an $SU(2)_L$ symmetry and an $SU(2)_R$ symmetry - the former, along with a contribution from the undotted quadratic term, represent the electroweak symmetry \cite{Raj, Kaushik}. The latter, along with a contribution from the dotted quadratic term, represents a $SU(2)_R \times U(1)_{grav}$ symmetry which is the  precursor of general relativity modified by a $U(1)_{grav}$. This symmetry is the right-handed counterpart of electroweak and is possibly a renormalisable theory - this might help us understand why general relativity is not renormalisable (it being a broken symmetry like the  weak force), whereas the $U(1)_{grav}$ is possibly the theoretical origin of MOND.

The imaginary corrections arise from multiplying an octonion onto itself; when they are significant, they might help understand why in the macroscopic limit 
space-time becomes classical. Because these corrections contribute an anti-self-adjoint part to the trace Hamiltonian. Whereas the Euclidean sector has no imaginary terms, is responsible for the strong force and for the newly proposed $SU(3)_{grav}$; it remains quantum and moreover does not take part in the cosmological expansion of space-time.  Also, it is evident that the weak force is a space-time symmetry, not an internal symmetry,  unlike the strong force. Together, gravitation and the weak force are broken symmetries in a 6D space-time, related to $SU(2)_L \times U(1)_Y \times SU(2)_R \times U(1)_{grav}$, and  stemming from the group-theoretic relation $SL(2,H) \sim SO(1,5)$. It can be argued that the two additional spatial dimensions here have a thickness of the order $(L_P/L)^{1/3}\sim 10^{-13}$ cm, where $L\sim 10^{28}$ cm is the size of the observed universe. This is not too far off from the range of the weak force, and also explains why we land up clubbing the weak force with the strong force as an internal symmetry, and not along with gravitation as a spacetime symmetry. If we were to club the four additional spatial directions (strong force) with the 6D spacetime, we have in effect a 10D space-time, motivated also by the group theory relation $SL(2,O)\sim SO(1,9)$. This 10D space-time evolves in Connes time $\tau$.

\section{Octonions, Clifford algebras, and elementary particles}
Spinors can be defined as minimal left ideals of Clifford algebras. Furthermore, one can use division algebras to construct Clifford algebras. And in specific cases, the symmetry properties of the corresponding spinors coincide with those of elementary fermions of the standard model. This gives strong evidence that complex quaternions and complex octonions are the natural home for defining states of quarks and leptons of three generations. Hence, not only do the octonions serve to define the coordinate system on the non-commutative manifold, but they also serve to naturally define states of quarks, leptons and gauge bosons, including those for gravitation. The standard model group symmetries do not have to be imposed by hand on these states; rather these symmetries are already present as subgroups in the symmetries of the octonion algebra. The five exceptional Lie groups - $G_2, F_4, E_6, E_7, E_8$ - all associated with the octonions, play a very important role in the deduction of the standard model. The all-encompassing role of octonions in defining non-commutative space-time, internal geometry, and particle states should be contrasted with the situation in quantum field theory: complex numbers for quantum states, and real numbers for space-time.

In the present context, the most important Clifford algebras are $Cl(2), Cl(3), Cl(6)$ and $Cl(7)$. The algebra $Cl(2)$ is generated by complex quaternions, keeping one of the quaternionic imaginary directions fixed. Spinors made from $Cl(2)$ are the left handed and right handed Weyl spinors, and the associated symmetry is the Lorentz algebra $SL(2,C)$. Octonions being a non-associative algebra, do not generate a Clifford algebra. However, maps are associative; therefore octonionic maps generate a Clifford algebra. The exceptional Lie group $G_2$ is the automorphism group of the seven imaginary directions of the octonion. $G_2$ has two maximal subgroups, $SU(3)$ and $SO(4)$. The former is the element stabiliser group (i.e. the automorphism group when one of the imaginary octonionic directions is kept fixed), whereas the latter is the stabiliser group of the quaternions inside the octonions. Keeping one of the imaginaries fixed, complexified octonionic maps generate the Clifford algebra $Cl(6)$. Spinors made from this algebra (there being eight of them) obey an $SU(3)$ symmetry: two out of these eight states are singlets of $SU(3)$, three are anti-triplets and three are triplets. A number operator $N$ made from the generators of $Cl(6)$ and having a $U(1)$ symmetry has the eigenvalues $0$ and $3$ for the singlet states, eigenvalue $1$ for the three anti-triplet states, and $2$ for the triplet states. Defining   $Q=N/3$ as the electric charge operator, we conclude that the singlet states are the neutrino and the positron, the anti-triplet is the anti-down quark, and the triplet is the up quark. 
Anti-particles are defined simply by the complex conjugation of these states. The $SU(3)$ symmetry is identified with $SU(3)_{color}$ of QCD, and the $U(1)$ symmetry with electromagnetism, $U(1)_{em}$. This inference is fully supported by the analysis of the fundamental Lagrangian in (\ref{aikyacn2}), as shown in \cite{Raj}. The Clifford algebra $Cl(6)$ describes one generation of standard model quarks and leptons under the unbroken symmetry $SU(3)_c \times U(1)_{em}$ \cite{Furey1, Furey2} . As noted earlier, this is an internal symmetry on Euclidean space (the vector bundle) and hence remains unbroken; whereas the weak interaction and general relativity, being space-time symmetries, are broken, as a result of the quantum-to-classical transition.

Among Clifford algebras, the algebras $Cl(3)$ and $Cl(7)$ are very special. They are the only ones, upto Bott periodicity, that have two irreducible representations (called pinors). The algebra $Cl(3)$, made from complex quaternions using all three imaginary quaternionic directions, is the algebra of complex split biquaternions. Each of the two quaternion copies corresponds to a $Cl(2)$ each, and one 
copy is the parity reverse of the other. This naturally enables the construction of one generation of chiral leptons and their anti-particles - left handed neutrino, right handed (sterile) neutrino, left handed electron and right handed electron, and their anti-particles \cite{Vaibhav}.

Analogously, the algebra $Cl(7)$ corresponds to complex split bioctonions and two copies of $Cl(6)$; thus describing one generation of chiral quarks and leptons. However, we introduce a significant subtlety: whereas the $U(1)_{em}$ associated with left-handed quarks and leptons has electric charge as its quantum number, the quantum number associated with the right-handed quarks and leptons is the square-root of mass (in Planck units). It takes the values $0, 1/3, 2/3, 1$ for the right-handed (sterile) neutrino, electron, up quark and down quark, respectively. We have switched the position of the electron and the down quark; the former is now a triplet, and the latter is a singlet, of the symmetry $SU(3)_{grav}$ associated with the right handed electron and the right handed up quark. Since the $SU(3)$ symmetry is unbroken, being an internal symmetry, the interaction $SU(3)_{color} \times SU(3)_{grav}$ is not parity violating. This might pave the way for resolving the strong CP problem, if we assume that  $SU(3)_{grav}$ is much weaker than $SU(3)_{color}$.

In this theory, there is evidence that there are three, and only three, fermion generations. This evidence comes from the triality of the group $SO(8)$, from the exceptional Jordan algebra $J_3(8)$, and from the symmetry $SU(3)_{gen}$ which arises in the branching of $E_8 \times E_8$ as we will see in the next section. The octonionic spinor states for the second and third generation can be obtained by applying $SU(3)$ rotations on the states of the first generation \cite{Singhfsc, Vivan}. It remains to be understood why there is only one copy of the gauge bosons, as opposed to  three.

\section{ $E_8 \times E_8$ unification of the standard model and pre-gravitation}
We propose that there are six fundamental forces (not four), and that they are described by the following symmetry
\begin{equation}
SU(3)_C \otimes \; SU(2)_{L} \otimes \; U(1)_Y \otimes \; SU(3)_{grav} \otimes\; SU(2)_{R} \otimes\; U(1)_{g}
\end{equation}
The first three of these are the standard model forces, and each of them has a gravitational counterpart, shown by the last three groups, which includes $SU(2)_R$, the precursor of general relativity. $SU(3)_{grav}$ and $U(1)_g$ are newly predicted. Whereas the sources for the standard model forces are color, weak isospin and hypercharge, the corresponding gravitational sources are gravi-color 
(non-zero for right-handed electron and right-handed up quark), gravi-isospin and gravi-hypercharge. Gravi-hypercharge is related to square-root mass in precisely the same way in which hypercharge is related to electric charge.

The origin of these forces lies in a specific branching of $E_8 \times E_8$. Thus the rep $(248,1)\oplus (1,248)$ is broken into two separate $E_8$. Each $E_8$ then branches as $SU(3)\otimes E_6$. This $SU(3)$ is mapped to an 8D vector space, which is identified with an octonion in case of the first $E_8$, and with the split part of a bioctonion in case of the second $E_8$. Together, the tensor product $SU(3)\otimes SU(3)$ maps to the split bioctonion (\ref{spbioc}) which in turn gives rise to the 
line-element shown in (\ref{thirteen}). Thus the branching brings with it the sought for unification of the vector bundle with space-time. 

$E_6$ is the only exceptional Lie group which has complex representations, and the two $E_6$ between them define three fermion generations, standard model gauge bosons, pre-gravitation, and two Higgs. Each $E_6$ branches as
\begin{equation}
E_6\rightarrow SU(3) \otimes SU(3) \otimes SU(3)
\end{equation}
and we have
\begin{equation}
248 = (8,1) \oplus (1,78) \oplus (3,27) \oplus (\bar{3}, \overline{27})
\end{equation}
One of the $SU(3)$ branches as $SU(2)\times U(1)$ and the two $E_6$ are interpreted as
\begin{equation}
{\rm First\ } E_6: SU(3)_{LHgen} \otimes SU(3)_c \times SU(2)_L \times U(1)_{Y}
\end{equation}
\begin{equation}
{\rm Second\ } E_6: SU(3)_{RHgen} \otimes SU(3)_{grav} \times SU(2)_R \times U(1)_{g}
\end{equation}
We get three generations of chiral fermions; for the left-handed ones the $U(1)$ quantum number is electric charge, and for the right handed ones the $U(1)$ quantum number is square root mass. Both electric charge as well as square-root mass are emergent entities, arising after the left-right symmetry breaking. The standard model Higgs gives mass to the left-handed fermions, and the newly predicted Higgs gives electric charge to the right-handed fermions. For further details of this unification proposal, the reader is referred to \cite{Kaushik}. 

This symmetry breaking, which is also the same as electroweak symmetry breaking, is enabled by the quantum-to-classical transition, in the very early universe as well as in the present low-energy universe. This transition happens when a very large number of fermions get entangled: for this to happen in the very early universe the universe must cool below a critical temperature (in this case the EW scale). Thus the role of the energy scale is only indirect; critical entanglement is what is actually responsible for symmetry breaking. Classical space-time emerges along with the dominance of classical objects, whose gravitation obeys the laws of general relativity in the vicinity of compact objects, and obeys the $U(1)_{grav}$ interaction in regions where gravitational acceleration is below the MOND critical acceleration. In the present context,  symmetry breaking implies that the broken $SU(2)_R$ symmetry becomes classical (general relativity) whereas the broken $SU(2)_L$ symmetry, i.e. the weak force, remains quantum in nature, and is short range. Even in today's universe, it is the act of quantum measurement (wave function collapse followed by the quantum-to-classical transition) which is responsible for the breaking of left-right symmetry. A quantum system, even at low energies, lives in its own split bioctonionic space.

The left-right symmetry breaking in the very early universe is also responsible for the separation of matter (has positive sign for square-root mass: $+\sqrt{m}$) and anti-matter (has negative sign for square-root mass: $-\sqrt{m}$). Like sign square-root masses attract pre-gravitationally, whereas unlike sign square-root masses  repel pre-gravitationally. This results in the separation of matter from 
anti-matter while preserving CPT symmetry of the matter-antimatter mirror universe, an idea which has been independently proposed earlier in \cite{Turok}. Pre-gravitation described by the $SU(2)_R$ symmetry is mediated by spin one gauge bosons, just like the standard model forces. After separation of matter and anti-matter in the very early universe, gravitation in our universe (fundamental observables still being the eigenvalues of the Dirac operator) appears as an `attractive only' interaction. This of course suggests that the quantum of gravitation is spin-2: however, a spin-2 graviton, even if it exists, cannot be fundamental but only composite.

Prior to the breaking of the $E_8 \times E_8$ symmetry, the dotted and undotted dynamical variables are unified into one, and the fundamental action (\ref{aikyacn2}) takes the form
\begin{equation}
\frac{S}{\hbar} = \frac{1}{2}\int \frac{d\tau}{\tau_{p}}\; Tr \biggl[\biggr. \dfrac{L_{p}^{2}}{L^{2}}  \dot{Q}_{1bioct}^\dagger\;  \dot {Q}_{2bioct} \biggr]
\label{deepacn}
\end{equation}
where
\begin{equation}
\dot{{Q}}_{1bioct} ^\dagger   =   \dot{{Q}}_{B}^\dagger + \dfrac{L_{p}^{2}}{L^{2}} \beta_{1} \dot{{Q}}_{F}^\dagger  ; \ \qquad \dot {{Q}}_{2bioct} =  \dot{{Q}}_{B} + \dfrac{L_{p}^{2}}{L^{2}} \beta_{2} \dot{{Q}}_{F}
\end{equation}
and
\begin{equation}
{\dot{{Q}}_B} = \frac{1}{L} (i\alpha q_B + L \dot{q}_B); \qquad  {\dot{{Q}}_F} = \frac{1}{L} (i\alpha q_F + L \dot{q}_F);
\end{equation}
The pre-gravitation sector (dotted variables) and the standard model sector (undotted variables) are unified, and so are the right handed fermions (dotted) and the left handed fermions (undotted). The bosonic and fermionic sectors are unified as well, and the only parameter in the action (\ref{deepacn}) is the dimensionless area $L_p^2/L^2$ (a scale-invariant theory). The trace Lagrangian is invariant under unitary transformations generated by the generators of $E_8 \times E_8$. The Yang-Mills coupling constant $\alpha$ arises after symmetry breaking, and because it appears as the relative weight between states of left handed fermions and states of right handed fermions, its value is determined by the octonion algebra. Similarly the value of $L_p/L$ is also determined by the octonion algebra, and there are no free parameters at all in the action (\ref{deepacn}).  The two dynamical variables in this action are functions only of Connes time, since the manifold (and the accompanying coordinate labels) emerge only after symmetry breaking.

This action can be thought of as describing a 2-brane with area $L_p^2 / L^2$.  The theory of interactions between 2-branes remains to be developed: interactions will be described as `collisions and scattering' in phase space, between matrix-valued dynamical variables.

We have at hand a matrix-valued Lagrangian dynamics on split bioctonionic space, which is the analog of the Heisenberg formulation of quantum dynamics. Dynamical variables evolve in (Connes) time, whereas states defined using the split bioctonions are time-independent.

\section{Applications: deriving the free parameters of the standard model}
From the Lagrange equations of motion, it is possible to deduce that the three fermion generations for a given value of the electric charge obey the Dirac equation in 10D spacetime. This equation is the eigenvalue equation for the (complexified) exceptional Jordan algebra $J_3(8)_C$ which has the symmetry group $E_6$. Here, $J_3(8)$, known as the exceptional Jordan algebra, comprises of $3\times 3$ Hermitean matrices with octonionic entries, and its automorphism group is $F_4$ \cite{Ramond, Dray, Manogue}. The characteristic equation of this algebra, a cubic equation with real eigenvalues, is of great interest as its eigenvalues uniquely determine the values of several fundamental constants of the standard model.

Consider a matrix of this algebra whose off-diagonal entries are the octonionic states representing fermion states of three generations with a given electric charge [i.e. one out of  $(0, 1/3, 2/3, 1)$]. Find the eigenvalues and eigenmatrices of this matrix, using the cubic characteristic equation. The eigenvalues, which are invariants (given the electric charge) and which we call Jordan eigenvalues, are shown in Fig. 4.
\begin{figure}[h]
\centering
\includegraphics[width=15cm]{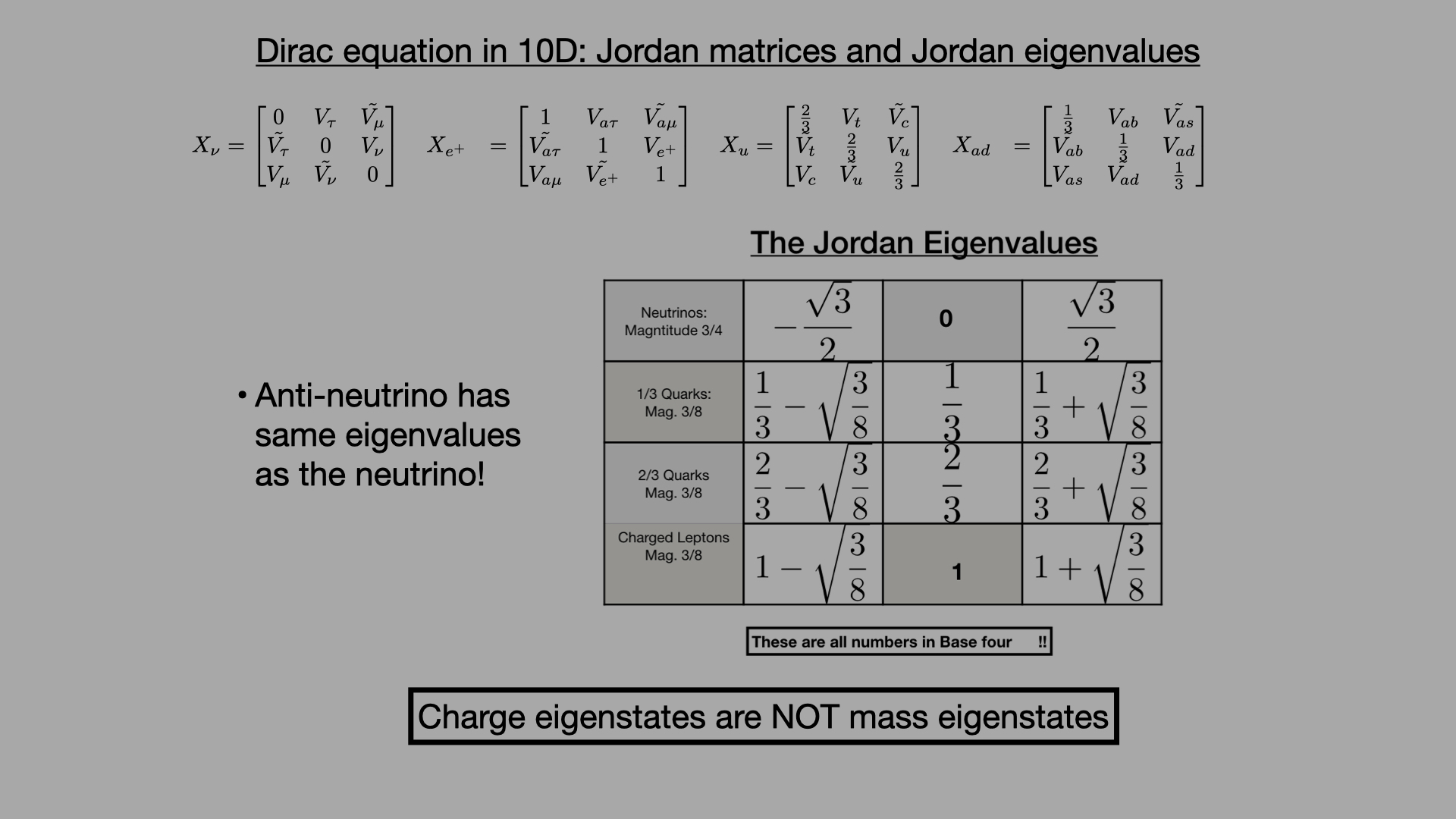}
\caption{The Jordan eigenvalues for the neutrino family, down quark family, up quark family, and electron family.}
\end{figure}
These twelve numbers are of great importance; the observed mass-ratios, as well as the low energy fine structure constant $1/137$ are derived from them. The eigenvalues for the anti-particles are opposite in sign to the values for their corresponding particles.

The octonionic states which have been employed here are charge eigenstates, which define the left-handed fermions. However, charge eigenstates are not mass eigenstates: the latter define the right-handed fermions, where after switching the position of the down quark and the electron the associated $U(1)$ quantum number takes the values $(0, 1/3, 2/3, 1)$ for square-root mass. Moreover, it takes the same values for all three generations, just as the electric charge takes the same values for all three generations. Why then do mass ratios appear so weird?! Note though that the first generation (square-root) mass ratios are very simple, being the same set as the charge ratios but with position of electron and down-quark interchanged. Only the second and third generation mass ratios are weird. If we calculate the Jordan eigenvalues for square-root mass eigenstates we get the same numbers as shown in Fig. 4 except that the position of the down quark family and that of the charged lepton family gets interchanged.

What exactly is the interpretation of the Jordan eigenvalues and where do the weird mass ratios come from? Interpretation of these eigenvalues: prior to left-right symmetry breaking the relative weight of the left handed states and right handed states (and hence the to-be-Yang-Mills coupling constant) is simply unity, and we can express this as
\begin{equation}
1 = \exp [0] = \exp [ Q + \sqrt{M} ]
\label{qM}
\end{equation}
From examining the Jordan eigenmatrices corresponding to the Jordan eigenvalues we conclude that the solution of this eigenvalue problem expresses charge eigenstates (or square-root mass eigenstates) as superpositions of eigenstates of `charge-square-root-mass', these being eigenstates of a $U(1)$ quantum number which necessarily takes the value $1/3$ (the value is $1/3$ for each generation, and $1/3 + 1/3 + 1/3 = 1$). This then is the interpretation of the Jordan eigenvalues: if we solve the eigenvalue problem for the Dirac equation in 10D spacetime for a given value of the electric charge, or for a given value of square-root mass, the three solutions are eigenmatrices of charge-square-root mass and the Jordan eigenvalues are the corresponding eigenvalues.

This permits eigenstates of square-root mass to be written as a superposition of eigenstates of electric charge. And since all our particle physics measurements are in the end based on the electromagnetic interaction (and not on the gravitational interaction) the mass ratios are determined by the relative weights of the charge eigenstates in this superposition, and the relative weights are in turn ratios of Jordan eigenvalues. This leads to a derivation of the strange observed mass ratios, which are in fact simple fractions as shown in Fig. 5. Had we been making our laboratory measurements using square-root-mass eigenstates, the mass ratios would have come out $(0, 1/3, 2/3, 1)$ for each of the three generations and the electric charge ratios would have been weird!
\begin{figure}[h]
\centering
\includegraphics[width=15cm]{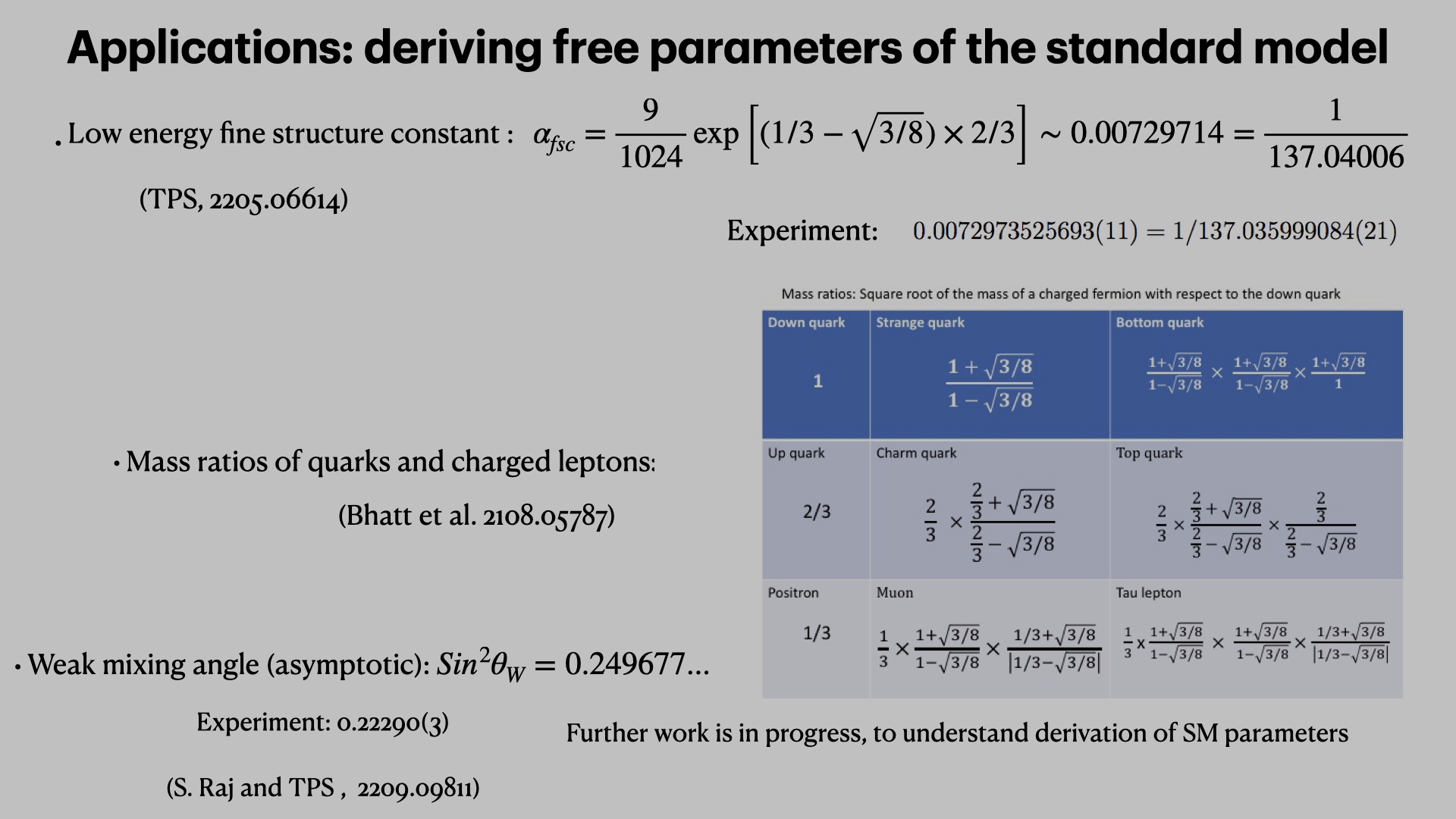}
\caption{Standard model parameters from eigenvalues of the exceptional Jordan algebra.}
\end{figure}

The theoretically calculated mass ratios match with observed values only if the neutrino is assumed to be Majorana. That is because the Jordan eigenvalues come out to be different when the neutrino is assumed to be a Dirac particle, and these values give drastically wrong ratios which do not agree with experiment. On the other hand, the so-called Koide ratio comes out to be exactly $2/3$ if the neutrino is assumed Dirac, and the ratio departs slightly from $2/3$, as is known from data, if the neutrino is assumed Majorana \cite{Singhep}. This gives support to our claim that only after wave function collapse the left-right symmetry is broken. The neutrino is fundamentally a Dirac particle, but is interpreted as a left-handed Majorana neutrino when its weak interaction is measured; it would register as a right-handed sterile Majorana neutrino if its gravitational interaction could be measured.

From the Lagrangian in (\ref{aikyacn2}) the expression for the low-energy fine structure constant can be read off to be $e^2/\hbar c \equiv \alpha^2 \times (L_P/L)^4$. Here, the constant $\alpha$ is related to the $Q$ in (\ref{qM}) by $\ln \alpha = Q$. A coefficient $\beta$ in front of the dotted part in ${\dot{{Q}}_B} = \frac{1}{L} (i\alpha q_B + L \dot{q}_B)$ is defined by $\ln\beta=\sqrt M$; with $\ln\alpha + \ln\beta = \ln \alpha\beta =Q+\sqrt M = 0 \implies \beta = 1/\alpha$. The constants $Q$ and $\sqrt{M}$ arise from the partition of the charge-square-root-mass quantum number $1/3$ into the weights $Q= (1/3 - \sqrt{3/8})/3$ and $\sqrt{M} = (\sqrt{3/8}-1)/3$ determined by the Jordan eigenvalues: $(1/3-\sqrt{3/8})$ for the left-handed down quark and $(\sqrt{3/8}-1)$ for its right handed counterpart, the electron.  This gives that $\alpha^2 = \exp [ 2/3 \times (1/3 - \sqrt{3/8})]$. From the construction of the octonionic states for the fermions, it can be concluded that \cite{Singhfsc} $(L_P/L)^4 = (\sqrt{1/32})^4$ and multiplying this by nine (because the square of the electric charge of the electron is nine times that of the down quark) gives the value of the low energy fine structure constant as shown in Fig. 5. Comparison with the experimental value is discussed in \cite{Singhfsc}. 

How do we know that this derivation is for the low energy limit? In fact this derivation is for the low interaction flat limit (\ref{thirteen}) in which the matrix-valued dynamical observables go to unity. We have then made an assumption (based on observations, and yet to be proved in this theory) that for electromagnetism, the low interaction limit is realised at low energies. The running of this constant with energy remains to be worked out, and for now we work with the running as derived using conventional methods of quantum field theory.

We note in passing that, analogous to the fine structure constant, the low energy $U(1)_{grav}$ gravitational fine structure constant $\alpha_g\equiv Gm_e^2/\hbar c$ is given by $\alpha_g^{1/2}=\beta^2 (L_P/L)^4$ and has the value
\begin{equation}
\alpha_g^{1/2} = \frac{9}{1024} \exp [ 2/3 \times (\sqrt{3/8} - 1/3) ] = 0.010586... = 1/94.4642...
\end{equation}
Therefore the gravitational fine structure constant $\alpha_g$ is
\begin{equation}
\alpha_g = 0.00011206378... = 1/8923.4892...
\end{equation}
The ratio of the strength of electromagnetic force to $U(1)_{grav}$ gravitational force is therefore
\begin{equation}
\alpha / \alpha_g = 65.11559...
\end{equation} 
This ratio is inevitably the primordial value, and is scaled down by an octonionic inflation which ends with the left-right symmetry breaking. This same inflation brings down the Higgs mass (as well as particle masses) from their Planck scale values to the presently observed values. These aspects are currently under investigation. 

We have also derived \cite{Raj} the weak mixing angle by investigating the bosonic part of the Lagrangian (\ref{aikyacn2}) and obtained the value shown in Fig. 5. In all, we have so far been able to derive ten standard model parameters: mass ratios, fine structure constant, and the weak mixing angle. If the octonionic theory is the correct theory of unification, it must also yield the remaining parameters from first principles: QCD coupling constant, quark mixing parameters (CKM matrix), neutrino mixing parameters (PMNS matrix) and Higgs mass. This is currently under investigation. We predict three right handed sterile neutrinos and this opens up the possibility that the observed neutrino oscillations are mediated by the sterile neutrinos, and that the neutrinos are in fact massless. Also, in the octonionic theory the cosmological constant is actually zero, and the role of dark energy is played by the uncollapsed atoms of space-ti me-matter: a preliminary attempt at showing this has been made in \cite{SinghDE}. 

\section{Left-right symmetry breaking: emergence of classical space-time and gravitation}
While we have remarked on this topic in earlier sections, it is important enough for us to reiterate some key points, as summarised in Fig. 6.
\begin{figure}[h]
\centering
\includegraphics[width=15cm]{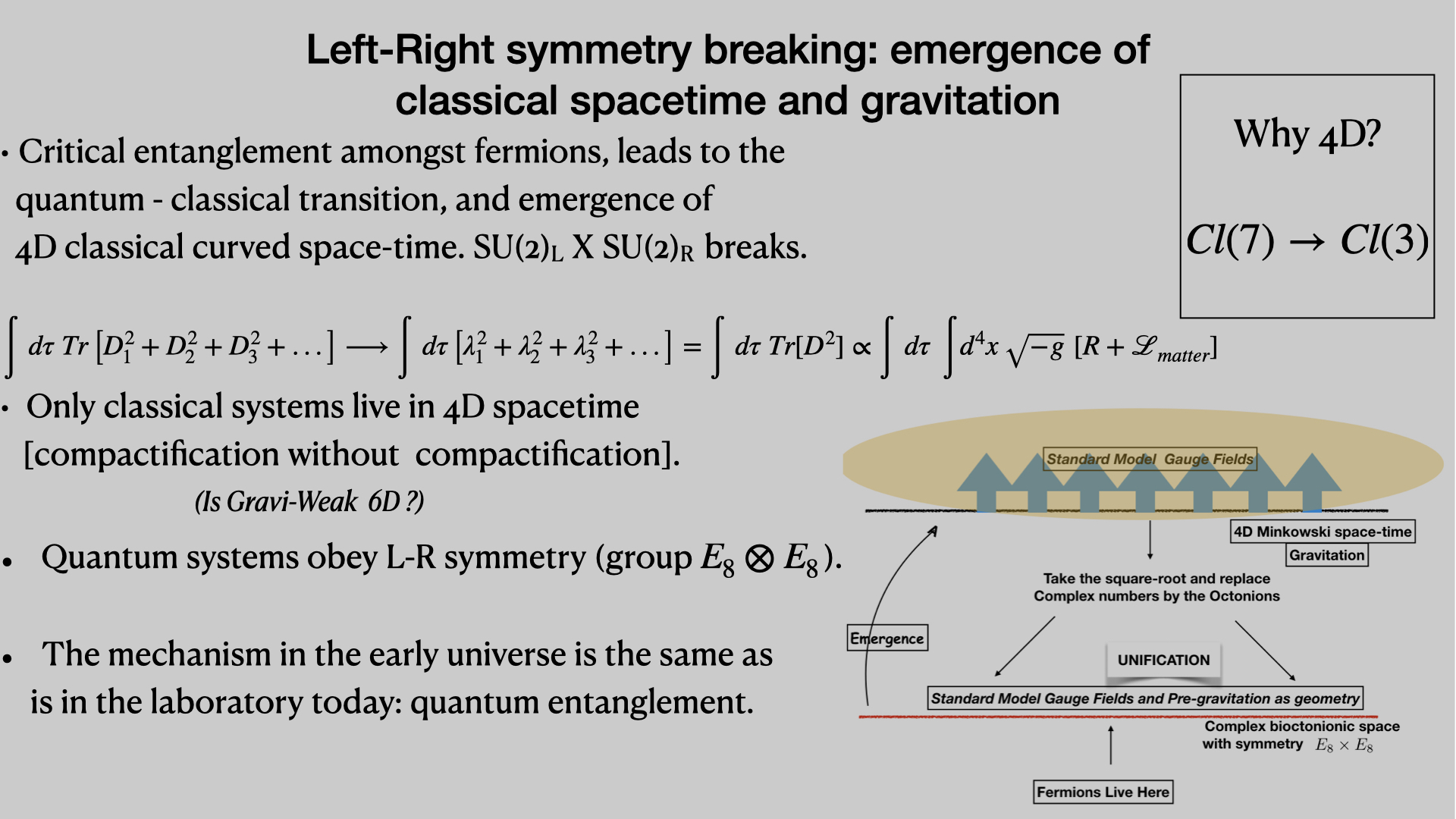}
\caption{Critical entanglement of fermions separates the emergent classical space-time from the internal symmetry of the strong interaction.}
\end{figure}
We have argued above that a line-element constructed from the split bioctonions in Eqn. (\ref{thirteen}) admits a 6D space-time, with an associated Lorentz symmetry $SO(1,5)$. Two of the spatial dimensions, whose geometry is responsible for the weak force, are much thinner than the other three, so that in the classical world we have a 4D spacetime with $SO(1,3)$ symmetry. The $SU(2)_R$ symmetry can be utilised to construct the equations of general relativity, in view of the spectral action principle. Emergence, initiated by spontaneous localisation, is the opposite of going from classical dynamics to trace dynamics. The Clifford algebra associated with the 6D space-time is $Cl(3)$ (related to complex split biquaternions), which in turn is made from two copies of $Cl(2)$. These account for chiral leptons - the neutrino and the electron, their anti-particles, and their second and third generation counterparts. When the strong interaction of quarks is included, one transits from $Cl(3)$ to $Cl(7)$ (complex split bioctonions) and to 10D spacetime. Quarks clearly cannot be confined to 6D or 4D spacetime: their strong interaction (this being the geometry of the four additional spatial directions) does not permit this, and this could be a possible explanation for quark confinement. On the other hand, all our measurements take place in 4D. Hence quarks only manifest in 4D spacetime through colorless composites such as protons and neutrons: not having color means the additional four spatial dimensions are not being probed by the composite state, but only by its constituents.

The point structure of spacetime is defined by the positions (the eigenvalues of the position operator to which collapse takes place) of the entangled fermions, these of course are the macroscopic classical objects of our universe. Since these positions commute, they impose a commutative point structure on the 4D spacetime. It helps to note that in objective collapse models, which is what is recovered from the octonionic theory, classical objects are nothing but short-lived quantum superpositions. This permits replacing the non-commutative algebra of quaternions by real numbers, which commute, thus enabling the transition to the classical 4D spacetime manifold to a great accuracy.

For a discussion of spin in trace dynamics see \cite{SinghSpin} and for proposal of a ground state in quantum gravity see \cite{Abhinash}.

\section{Are there any testable predictions?}
Figure 7 below lists predictions of the octonionic theory which could become testable with future development in technology.
\begin{figure}[h]
\centering
\includegraphics[width=15cm]{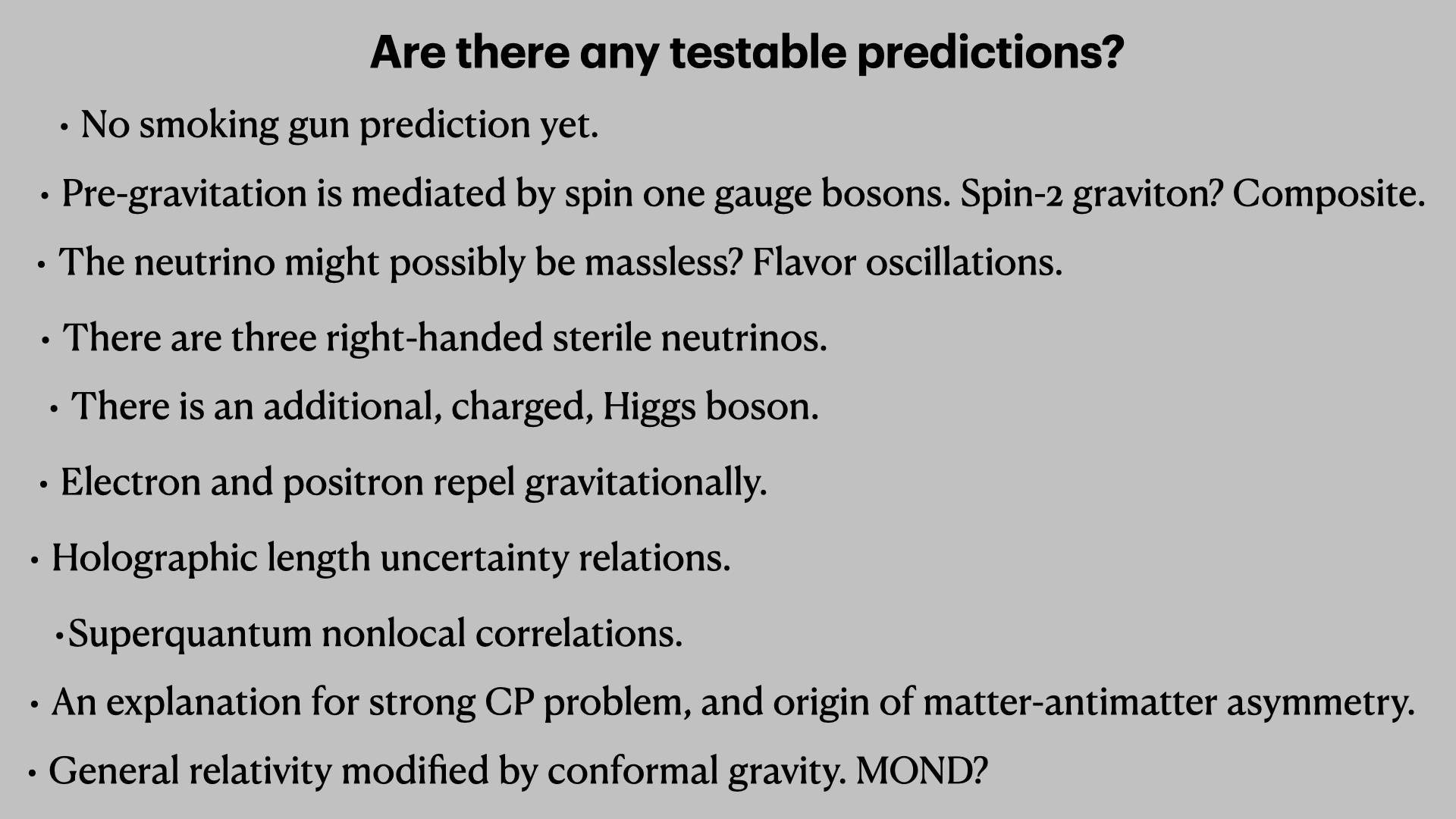}
\caption{Predictions of the octonionic theory.}
\end{figure}
There is no smoking gun prediction yet, where a measurable effect has quantitatively different predictions from quantum field theory, as compared to the prediction of the octonionic theory. There are postdictions; i.e. the theoretical derivation of the free parameters of the standard model. 

We believe we have provided adequate evidence that this is the correct path to quantum gravity and unification, and to a proper understanding of the standard model. The octonionic theory is entirely motivated by addressing a foundational problem of quantum theory: the assumption of classical spacetime in quantum theory is an approximation even at low energies; an assumption which needs to be dropped. And when it is dropped, doing so opens a path to quantum gravity and to unification, and to a first principles derivation of the experimentally measured constants of the standard model.

Undoubtedly, much remains to be done still, before one can claim to have a theory that fully explains current data of the standard model. The following are the key aspects that remain to be addressed, and are currently being investigated.

\begin{itemize}

\item SM parameters:  Higgs mass, W and Z masses, quark mixing matrix, neutrinos mixing matrix, QCD coupling constant.

\item The quantum-to-classical transition: spontaneous localisation.

\item  Consequences of working with a quaternionic / octonionic spinor spacetime: role in condensed matter systems?

\item Develop an EFT to take account of octonionic corrections to QFT.

\item Understanding the gravi-weak interaction. 

\item Implications for cosmology: are MOND and RMOND consequences of the octonionic theory?

\end{itemize}

\ack The proposed octonionic theory relies very significantly on important earlier work by previous researchers. These include theories of spontaneous collapse of the wave function, the theory of trace dynamics, the spectral action principle, and the very detailed and comprehensive research on applications of the octonions to particle physics. To all these researchers, I am deeply grateful. Without the foundations laid by them, the octonionic theory would simply not exist. I would like to thank my collaborators without whose contribution the developments reported here would not have been possible. In particular I wish to thank Vatsalya Vaibhav, Priyank Kaushik, Rabsan Galib Ahmed and Sherry Raj for our collaborative work during the last two years, which has brought the octonionic theory to its present robust and believable shape. It is a pleasure to thank Felix Finster, Jos\'e M. Isidro, Claudio Paganini, Cenalo Vaz, Kinjalk Lochan, Sukratu Barve, Kartik Kakade, Avnish Singh and Aditya Ankur Patel for our ongoing present collaboration. I would like to thank the participants of DICE2022 for stimulating discussions, and the organisers for providing a vibrant platform for exchanging new ideas and results.

A note on the references: the list of references below is far from being exhaustive. By and large only those works have been listed which directly impact the research reviewed in this article. In particular, the very large body of research on octonions in particle physics is not explicitly referred to; for a comprehensive listing please see \cite{Singh2021}.

\section*{References}

 

\end{document}